\documentclass{article}
\usepackage{spconf,amsmath,graphicx}
\usepackage{multirow}       
\usepackage{array,booktabs}
\usepackage{hyperref}
\usepackage{amsfonts}       

\usepackage{subcaption}
\usepackage{amsthm}
\usepackage{amsmath}

\def\x{{\mathbf x}}

\title{SEED: Sound Event Early Detection via Evidential Uncertainty}
\name{Xujiang Zhao$^{\dagger}$, Xuchao Zhang$^{\ddagger}$, Wei Cheng$^{\ddagger}$, Wenchao Yu$^{\ddagger}$, Yuncong Chen$^{\ddagger}$, Haifeng Chen$^{\ddagger}$, Feng Chen$^{\dagger}$}
\address{$^{\dagger}$University of Texas at Dallas, \texttt{\{xujiang.zhao,feng.chen\}@utdallas.edu} \\ $^{\ddagger}$NEC Laboratories America, \texttt{\{xuczhang,weicheng,wyu,yuncong,haifeng\}@nec-labs.com}
}

\begin{document}
%
\maketitle
\begin{abstract}
Sound Event Early Detection (SEED) is an essential task in recognizing the acoustic environments and soundscapes.
However, most of the existing methods focus on the offline sound event detection, which suffers from the over-confidence issue of early-stage event detection and usually yield unreliable results. To solve the problem, we propose a novel Polyphonic Evidential Neural Network (PENet) to model the evidential uncertainty of the class probability with Beta distribution. Specifically, we use a Beta distribution to model the distribution of class probabilities, and the evidential uncertainty enriches uncertainty representation with evidence information, which plays a central role in reliable prediction.
To further improve the event detection performance, we design the backtrack inference method that utilizes both the forward and backward audio features of an ongoing event. 
Experiments on the DESED database show that the proposed method can simultaneously improve 13.0\% and 3.8\% in time delay and detection F1 score compared to the state-of-the-art methods.
\end{abstract}
\vspace{-3mm}
\section{Introduction}\label{sec:intro}
\vspace{-2mm}
Sound event detection (SED), as a fundamental task to recognize the acoustic events, has achieved significant progress in a variety of  applications, such as unobtrusive monitoring in health care, and surveillance.
Recently, Deep Neural Network (DNN) based methods such as CRNN~\cite{mesaros2019sound} and Conformer~\cite{miyazaki2020conformer} significantly improve the event detection performance. However, these methods are usually designed in an offline setting that the entire audio clip containing sound events is fully observed. This assumption may not hold in many real-world applications that require real-time event detection. For example, the event detection in audio surveillance~\cite{viet2013real} requires low latency reaction to potentially dangerous circumstances for life saving and protection.
In this paper, we will focus on the sound event early detection (SEED) problem, which is designed in an online setting that requires ongoing events to be recognized as early as possible.

\begin{figure}[!t]
    \centering
    \begin{subfigure}[b]{0.23\textwidth}
        \centering
        \includegraphics[width=\linewidth]{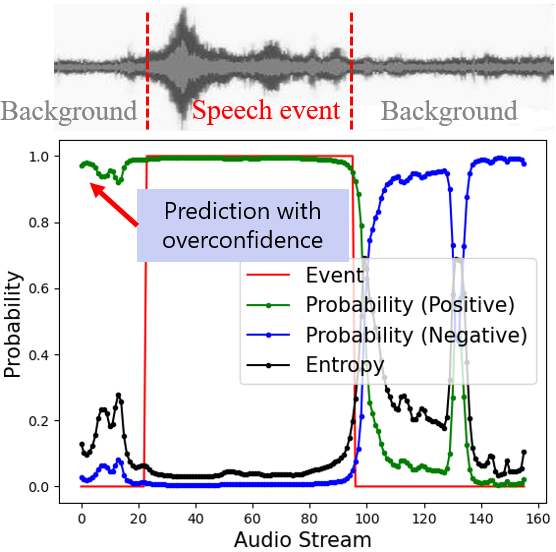}
        \caption{Baseline}
        \label{fig1a}
    \end{subfigure}
    \begin{subfigure}[b]{0.23\textwidth}
        \centering
        \includegraphics[width=\linewidth]{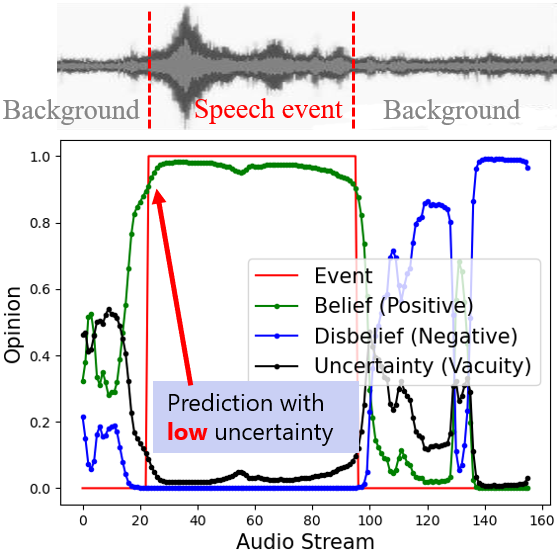}
        \caption{Ours}
         \label{fig1b}
    \end{subfigure}
    \vspace{-2mm}
    \small{
    \caption{An example of sound event early detection. (a) Baseline (CRNN)  model might give a prediction with over-confidence, and result in a false positive detection; (b) To avoid over-confidence, our framework estimates vacuity uncertainty instead of entropy, and make a reliable prediction with low vacuity.}
    \label{fig:example}
    \vspace{-6mm}
    }
\end{figure}

Despite the importance of the SEED problem, few existing focus on detecting sound events with short delays from audio streaming data. 
Some works design a monotonous detection function to achieve early detection, such as random regression forests algorithm\cite{phan2015early}, Dual-DNN~\cite{phan2018enabling}.
Some work ~\cite{mcloughlin2018early} proposes a detection front-end to identify seed regions from spectrogram features to detect events at the early stage.
However, the prediction of these methods are based on probability, which could be not reliable (over-confidence)~\cite{zhao2020uncertainty, sensoy2018evidential}. Especially during the early stage of an ongoing event, we only collect a small number of stream audios 
that may not be enough to compose a clear event sound to support a reliable prediction. Figure~\ref{fig:example} (a) shows an example that prediction based on probability is over-confidence at the early stage.

To solve the issue discussed above, we propose a novel Polyphonic Evidential Neural Network (PENet) to estimate the Beta distribution instead of the class probability such that we can estimate evidential uncertainty for each prediction. The attached evidential uncertainty is able to detect the ``over-confidence'' prediction and achieve a reliable prediction. To further improve the SEED performance, we propose the backtrack inference method that consider the forward information (waiting for the future information) of an ongoing event. 

\vspace{-2mm}
\section{Methodology}\label{sec:method}
\vspace{-2mm}
In this section, we begin with the essential concepts of evidential uncertainty. Then, we introduce the proposed Polyphonic Evidential Neural Network with its backtrack inference method.

\vspace{-2mm}
\subsection{Subjective Logic and Evidential Uncertainty}\label{sec:SL}
\textbf{Subjective Logic} (SL)~\cite{josang2016subjective} defines a subjective opinion by explicitly considering the dimension of uncertainty derived from vacuity (i.e., a lack of evidence). For a given binomial opinion towards proposition (e.g., an audio segment) $\x$, an opinion is expressed by two belief masses (i.e., belief $b$ and disbelief $d$) and one uncertainty mass (i.e., vacuity, $u$). Denote an opinion by $\omega= (b, d, u)$,
where $b$ and $d$ can be thought as positive (event happen) vs. negative (event not happen) on a given audio segment. We have the property $b+d+u=1$ and $b, d, u \in [0, 1]$.

An opinion, $\omega$, can always be projected onto a single probability distribution by removing the uncertainty mass. To this end, the expected belief probability $p$ is defined by: $p = b+a\cdot u$, where $a$ refers to a base rate representing a prior knowledge without commitment such as neither agree nor disagree.
A binomial opinion follows a Beta pdf (probability density function), denoted by $\text{Beta}(p | \alpha, \beta)$ in Eq~\eqref{beta_pdf}, where $\alpha$ and $\beta$ represents the positive and negative evidence.
\begin{eqnarray} \label{beta_pdf}
\vspace{-1mm}
\textbf{Beta}(p | \alpha, \beta) = \frac{1}{B(\alpha, \beta)}p^{\alpha-1}(1-p)^{\beta-1},
\end{eqnarray} 
where $B(\alpha, \beta) = \Gamma(\alpha)\Gamma(\beta)/ \Gamma(\alpha+\beta)$ and $\Gamma(\cdot)$ is the gamma function.
In SL, $\alpha$ and $\beta$ are received over time. An opinion $w$ can be obtained based on $\alpha$ and $\beta$ as $w = (\alpha, \beta)$. This can be translated to $w = (b, d, u)$ using the mapping rule in SL:
\begin{eqnarray}
\vspace{-2mm}
b = \frac{\alpha-1}{\alpha +\beta }, \; d = \frac{\beta-1}{\alpha +\beta}, \; u = \frac{W}{\alpha +\beta},
\label{eq:beta-mapping-rule}
\end{eqnarray} 
where $W$ is an amount of uncertainty evidence. In practice we set $W=2$ for binary case.

\noindent {\bf Evidential Uncertainty.} The concept of evidential uncertainty has been discussed differently depending on domains~\cite{josang2018uncertainty, zhao2020uncertainty,xu-etal-2021-boosting,shi2020multifaceted,hu2020multidimensional}. In this work, we adopt the concept of uncertainty based on SL in developing an uncertainty-based SEED framework when the input is a streaming audio signal. 
{\em Vacuity} refers to a lack of evidence, meaning that uncertainty is introduced because of no or insufficient information. High vacuity might happen at the early stage of an ongoing event due to the small amount of collected stream audios, resulting in an over-confidence estimation. Table~\ref{tab:evidence} illustrates the difference between probability and evidence. For example, at the early stage of an ongoing event, we only collect 1 positive evidence and 4 negative evidence. And we can calculate its expected probability $p=[0.2, 0.8]$, which result in high confidence of negative prediction. However, prediction based on small amount of evidence (i.e., high vacuity) is not reliable. With collecting more evidence (e.g., $[\alpha, \beta]=[200, 4]$), we would have a reliable prediction with low vacuity.

\begin{table}[ht!]
\footnotesize
\centering
\vspace{-1mm}
\caption{Difference between evidence and probability. Prediction with less evidence (high vacuity) is not reliable.}
\vspace{-2mm}
\begin{tabular}{l|c|c|c}
 \toprule
\textbf{Evidence}  & $[\alpha, \beta]=[1, 4]$ & $[\alpha, \beta]=[4, 4]$ & $[\alpha, \beta]=[200, 4]$  \\ 
\midrule
\textbf{Probability} & $p=[0.2, 0.8$] & $p=[0.5, 0.5]$ & $p=[0.98, 0.02]$   \\ 
\midrule
\textbf{Uncertainty} &\textbf{High Vacuity} &\textbf{High Vacuity} & \textbf{Low}  \\ 
\bottomrule
\end{tabular}
\label{tab:evidence}
\vspace{-4mm}
\end{table}

\subsection{Polyphonic Evidential Neural Network}
\vspace{-1mm}

\begin{figure*}[!t]
    \centering
    \includegraphics[width=0.94\textwidth]{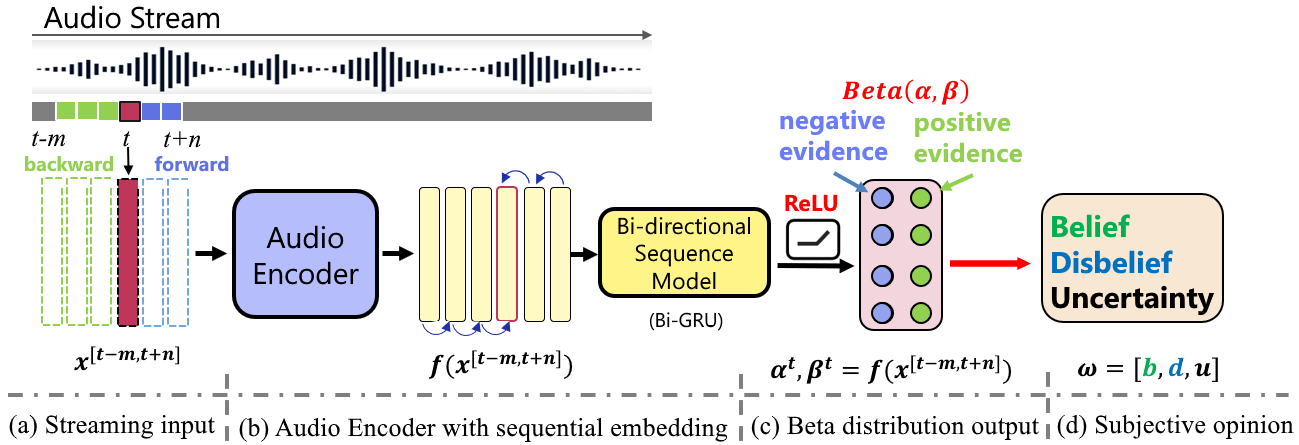}
    \caption{\textbf{PENet Overview.} Given the streaming audio segments (a), PENet is designed for estimating the Beta distribution (c), which can be transferred to subjective opinion (d) with vacuity uncertainty.}
    \label{fig:framework}
\end{figure*}
Based on the intuition of evidential uncertainty in SEED, we propose a novel Polyphonic Evidential Neural Network (PENet) with reliable prediction. The overall description of the
framework is shown in Figure~\ref{fig:framework}. For SEED setting, the audio signal is collected in a streaming way. At each timestamp $t$, we collect an audio segment $\x^t$. The corresponding label of $\x^t$ is denoted as $y^t=[y^t_1, \ldots, y^t_K]$, where $y^t_k=\{0, 1\}$.

\noindent {\bf PENet.}
For polyphonic sound event detection, most existing methods would consider a binary classification for each class, such as softmax output~\cite{Turpault2019_DCASE,hershey2021benefit}. 
As discussed in Section~\ref{sec:SL}, evidential uncertainty can be derived
from binomial opinions or equivalently Beta distributions to model  an event distribution for each class. Therefore, we design a Polyphonic Evidential Neural Network (PENet) $f$ to form their binomial opinions for the class-level Beta distribution of a given audio segment $\x^t$.  In addition, we considered a context of $m$ segments for sequential input purpose.
Then, the conditional probability $P(p^t_k|\x^{[t-m, t]};\theta)$ of class $k$ can be obtained by:
\begin{eqnarray}
\vspace{-2mm}
P(p^t_k|\x^{[t-m, t]};\theta) &=&\textbf{Beta}(p^t_k|\alpha^t_k, \beta^t_k) \\ 
\alpha_k^t, \beta_k^t &=& f_k(\x^{[t-m, t]};\theta) \label{eq:mlenn}
\vspace{-2mm}
\end{eqnarray}
where $\x^{[t-m, t]}$ means a sequence of audio segments, i.e., $[\x^{t-m}, \x^{t-m+1},\ldots, \x^t]$, $f_k$ is the output of PENet for class $k$, and $\theta$ is the model parameters. The Beta probability function $\text{Beta}(p_k^t|\alpha_k^t, \beta_k)$ is defined by Eq.~\eqref{beta_pdf}. 
Note that PENet is similar to the classical polyphonic sound event detection model (e.g., CRNN~\cite{Turpault2019_DCASE}), except that we use an activation layer (e.g., ReLU) instead of the softmax layer (only outputs class probabilities). This ensures that PENet would output non-negative values taken as the evidence for the predicted Beta distribution.

\noindent {\bf Training with Beta Loss.} In this paper, we design and train neural networks to form their binomial opinions for the classification of a given audio segment as a Beta distribution. For the binary cross-entropy loss, we have the Beta loss by computing its Bayes risk for the class predictor,
\begin{eqnarray}
\vspace{-2mm}
\mathcal{L}_{Beta} &=& \sum_{t=1}^T \sum_{k=1}^K \int \Big[\textbf{BCE}(y_k^t, p_k^t) \Big] \textbf{Beta}({p}_k^t;\alpha_k^t, \beta_k^t) d {p}_k^t \nonumber \\
&=& \sum_{t=1}^T \sum_{k=1}^K \Big[y_k^t\Big(\psi(\alpha_k^t+\beta_k^t) - \psi(\alpha_k^t)\Big) +
\nonumber \\
&&(1-y_k^t)\Big(\psi(\alpha_k^t+\beta_k^t) - \psi(\beta_k^t)\Big) \Big],
\label{eq:loss}
\end{eqnarray} 
where $T$ is the number of segments decomposed from an audio, $K$ is the number of class, $\textbf{BCE}(y_k^t, p_k^t)= -y_{k}^t \ln(p_{k}^t) - (1-y_k^t) \ln (1-p_{k}^t)$ is the binary cross-entropy loss, and $\psi(\cdot)$ is the \textit{digamma} function. The log expectation of Beta distribution derives the second equality.

\noindent {\bf Uncertainty-based Inference.}
At the test stage, we consider a simple strategy to make a reliable prediction. For each class, we predict sound events happened only when belief larger than disbelief with a small vacuity,
\vspace{-2mm}
\begin{eqnarray}
\hat{y}_k^t =\begin{cases}1,& \text{if } b^t_k > d^t_k \text{ and } u_k^t < V \\0 ,& \text{otherwise}\end{cases}
\label{eq:inference}
\end{eqnarray} 
where $\hat{y}_k^t\in \{0, 1\}$ is the model prediction for class $k$ in segment $t$, $V$ is the vacuity threshold.
\vspace{-2mm}
\subsection{Backtrack Inference}
\vspace{-2mm}
We propose a backtrack inference method that considers forward and backward information to feed into PENet as a sequential input to further improve early detection performance. Figure~\ref{fig:framework} (a) illustrate the backtrack input. Then, we can rewrite Eq.~\eqref{eq:mlenn} as
\begin{eqnarray}
\alpha_k^t, \beta_k^t = f_k(\x^{[t-m, t+n]};\theta),
\label{eq:mlenn2}
\end{eqnarray} 
where $m$ is the backward steps, and $n$ is the forward steps.
The additional previous and future information is critical for the prediction of the current audio segment. We show that backtrack inference improves SEED detection accuracy, but the waiting process (consider forward information) causes a higher detection delay.

\section{Experiments}\label{sec:experiment}

\subsection{Experiment Details}\label{sec:details}

\noindent {\bf Dataset.} We conduct the experiments on DESED2021 dataset~\cite{Turpault2019_DCASE}
The dataset for this task is composed of 10 sec audio clips recorded in domestic environments or synthesized using Scaper~\cite{salamon2017scaper} to simulate a domestic environment. The dataset includes 10 classes of sound events that represent a subset of Audioset~\cite{hershey2021benefit}. In DESED2021 dataset, the training set contains 10,000 synthetic audio clips with strong-label, 1578 weak-label audio clips, and 14,412 unlabeled audio clips. The validation set includes 1168 audio clips that are annotated with strong-label (timestamps obtained by human annotators).
The test set includes 1,016 real-world audio clips.

\noindent {\bf Features.} 
The input features used in the experiments are log-mel spectrograms extracted from the audio signal resampled to 16000 Hz.
The log-mel spectrogram uses 2048 STFT windows with a hop size
of 256 and 128 Mel-scale filters. At the training stage, the input is the full observed 10-second sound clip. As a result, each 10-second sound clip is transformed into a 2D time-frequency representation with a size of (626×128).
At the test stage, we collect an audio segment at each timestamp, which can be transformed into a 2D time-frequency representation with a size of (4×128).

\noindent {\bf Comparing Methods.}
To evaluate the effectiveness of our proposed approach (PENet), we compare it with one state-of-the-art SEED method: Dual DNN~\cite{phan2018enabling}; two SED methods: CRNN~\cite{Turpault2019_DCASE} and Conformer~\cite{miyazaki2020conformer}; and three different uncertainty methods: \textit{Entropy}, 
\textit{Epistemic} uncertainty~\cite{gal2016dropout} (represents the uncertainty of model parameters), and \textit{Aleatoric} uncertainty~\cite{depeweg2018decomposition} ( represents the uncertainty of data noise). We use MC-drop~\cite{gal1506bayesian} to estimate epistemic and aleatoric uncertainties in the experiments.

\noindent {\bf Evaluation Metrics.} Since the traditional offline sound event detection metrics cannot early detection performance, we use both early detection F1 and detection delay to evaluate our performance for the onset of sound events at the early stage. 
We first define the true positive prediction for the event $k$ only happened when the first prediction timestamp $d_p$ is located into event happened intervals. In addition, we set an early predict tolerance $L$ that if the first prediction is earlier than true event occurred. Otherwise, we consider the prediction for this event is false positive,
\begin{eqnarray}
\text{TP}_k =\begin{cases}1, & \text{if }y^{d_p}_k==1 \text{ and } d_p-d_t \ge L \\ 0 ,& \text{otherwise}  \end{cases}
\label{eq:tp}
\end{eqnarray} 
where $d_t$ is the onset timestamp of the predicted event. For detection delay, it's only measured when we have a true positive prediction. The detection delay is defined as follows,
\begin{eqnarray}
\text{delay} =\begin{cases}d_p-d_t, & \text{if } d_p\ge d_t \\ 0 ,& \text{if } d_p< d_t  \end{cases}
\label{eq:delay}
\end{eqnarray} 

\noindent {\bf Set up.} For all experiments, we use CRNN~\cite{turpault2020training} as the backbone except Conformer. We use the Adam optimizer for all methods and follow the same training setting as~\cite{turpault2020training}. For the uncertainty threshold, we set 0.5 for epistemic uncertainty and 0.9 for other uncertainties (entropy, vacuity, aleatoric).

\vspace{-2mm}
\subsection{Results and Analysis}\label{sec:result}

\noindent {\bf Early Detection Performance.}
Table~\ref{tab:results1} shows that our Evidence model with vacuity
outperforms all baseline models under the detection delay and early detection F1 score for sound event early detection. The
outperformance of vacuity-based detection is fairly impressive. This confirms that low vacuity (a large amount of evidence) is the key to maximize the performance of early detection. In addition, we observe that backtrack technical can significantly improve the early detection F1, demonstrating that backtrack information is essential in SEED. However, using the backtrack technical would increase the detection delay as well. Furthermore, the test inference time of our approach is around 5ms, less than the streaming segment duration (60ms), which indicates that our method satisfies the real-time requirement.

\begin{table}[ht!]
\centering
\caption{Early detection performance on DESED dataset.}
\begin{tabular}{l|c|c|c}
 \toprule
\textbf{Model} & \textbf{Delay} $\downarrow$ & \textbf{F1} $\uparrow$ & \textbf{Time}  \\ 
\midrule
Conformer & 0.372  & 0.639 & 6.6ms  \\
Dual DNN & 0.386  & 0.682 & 5.1ms  \\
CRNN & 0.284  & 0.687 & 5.0ms  \\
CRNN + entropy & 0.312  & 0.679& 5.0ms  \\
CRNN + epistemic & 0.278  &  0.647& 27ms \\
CRNN + aleatoric & 0.281  &  0.643& 27ms \\

\midrule
PENet & \textbf{0.247}  & 0.670& 5ms  \\
PENet + vacuity & 0.252  & 0.691& 5ms  \\
PENet + vacuity + backtrack & 0.310 & \textbf{0.725} & 5.2ms   \\
\bottomrule
\end{tabular}
\label{tab:results1}
\end{table}

\begin{figure}[!t]
    \centering
    \includegraphics[width=0.35\textwidth]{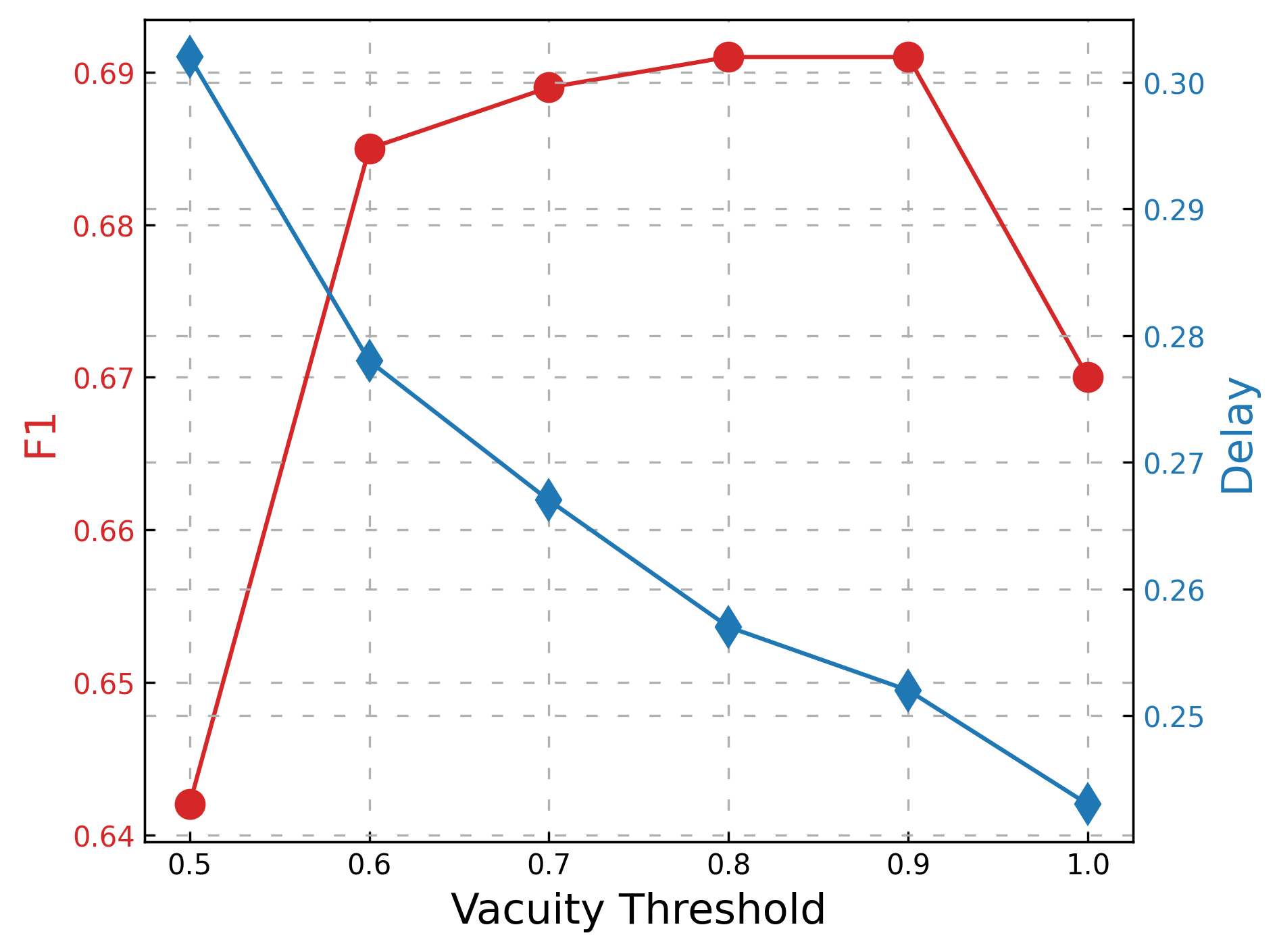}
    \caption{Effect of vacuity threshold.}
    \label{fig:vacuity}
\end{figure}

\noindent {\bf Uncertainty Analysis.} We explore the sensitivity of vacuity threshold used in the evidence model. Figure~\ref{fig:vacuity} plots the detection delay and early detection F1 score with the varying vacuity threshold values. When the vacuity threshold increases, the detection delay decreases continuously, and the early detection F1 score reaches the highest when vacuity is 0.9.

\begin{figure}[!t]
    \centering
    \includegraphics[width=0.35\textwidth]{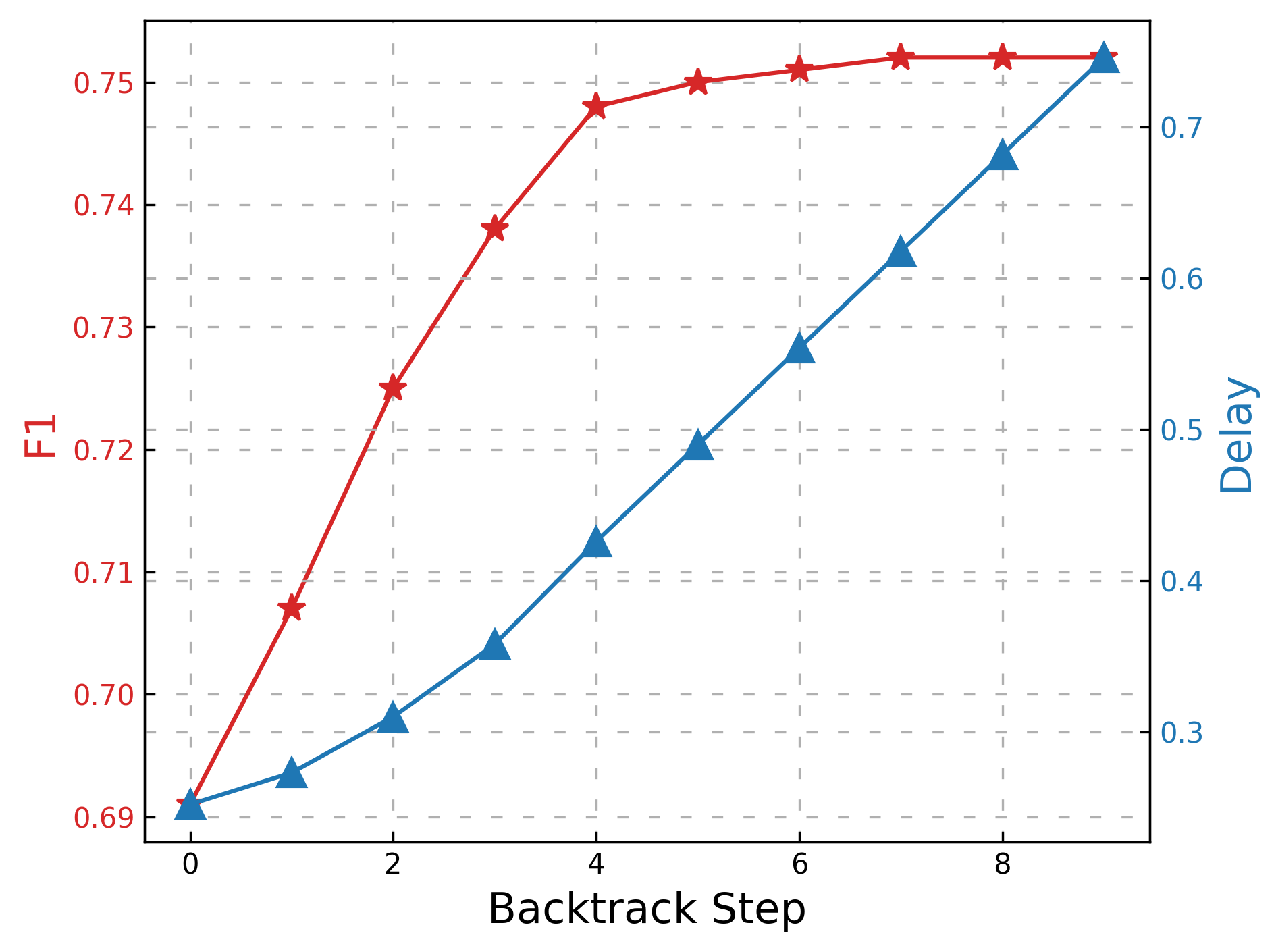}
    \caption{Detection delay and F1 score for different numbers of backtrack steps.}
    \label{fig:backtrack}
    \vspace{-3mm}
\end{figure}

\noindent {\bf Trade off of Backtrack.}
We analyzed the sensitivity of our proposed backtrack method to the number of backtrack steps. Table~\ref{fig:backtrack} shows a trade-off between detection delay and F1 score with the varying numbers of steps. When the backtrack step increase, the  detection delay is continuously increased, and detection F1 increases until backtrack step equal to 6.
The results demonstrate that backtrack information is critical to improving the detection accuracy in SEED. 



\vspace{-2mm}

\section{Conclusion}
In this paper, we propose a novel Polyphonic Evidential Neural Network to model the evidential uncertainty of the class probability with Beta distribution. Specifically, we use a Beta distribution to model the distribution of class probabilities, and the evidential uncertainty enriches uncertainty representation with evidence information, which plays a central role in reliable prediction.
And the proposed backtrack inference method can further improve the event detection performance.
The experiment results demonstrate that the proposed method outperformed in the SEED task compared with other competitive counterparts. 

\newpage
\bibliographystyle{IEEEbib}
\bibliography{main}

\end{document}